\pdfoutput=1
\documentclass[natbib]{svjour3}                     
\smartqed  
\usepackage{graphicx,amssymb}
\newcommand{\aap}{{Astron. Astrophys.}}
\newcommand{\apj}{{Astrophys. J.}}
\newcommand{\apjl}{{Astrophys. J. Lett.}}
\newcommand{\apjs}{{Astrophys. J. Supp.}}
\newcommand{\mnras}{{Mon. Not. Roy. Astron. Soc.}}
\newcommand{\prd}{{Phys. Rev. D}}

\begin{document}

\title{Current Status of Simulations
}


\author{P. Chris Fragile
}


\institute{P. C. Fragile \at
              Department of Physics \& Astronomy \\
              College of Charleston \\
              Charleston, SC 29424, USA \\
              Tel.: 843-953-3181\\
              Fax: 843-953-4824\\
              \email{fragilep@cofc.edu}           
}

\date{Received: date / Accepted: date}

\maketitle

\begin{abstract}
As the title suggests, the purpose of this chapter is to review the current status of numerical simulations of black hole accretion disks.  This chapter focuses exclusively on {\it global} simulations of the accretion process within a few tens of gravitational radii of the black hole.  Most of the simulations discussed are performed using general relativistic magnetohydrodynamic (MHD) schemes, although some mention is made of Newtonian radiation MHD simulations and smoothed particle hydrodynamics.  The goal is to convey some of the exciting work that has been going on in the past few years and provide some speculation on future directions.

\end{abstract}

\section{Introduction}
\label{sec:intro}

Going hand-in-glove with analytic models of accretion disks, discussed in Chapter 2.1, are direct numerical simulations.  Although analytic theories have been extremely successful at explaining many general observational properties of black hole accretion disks, numerical simulations have become an indispensable tool in advancing this field.  They allow one to explore the full, non-linear evolution of accretion disks from a first-principles perspective.  Because numerical simulations can be tuned to a variety of parameters, they serve as a sort of ``laboratory'' for astrophysics.

The last decade has been an exciting time for black hole accretion disk simulations, as the fidelity has become sufficient to make genuine comparisons between them and real observations.  The prospects are also good that within the next decade, we will be able to include the full physics (gravity + hydrodynamics + magnetic fields + radiation) within these simulations, which will yield complete and accurate numerical representations of the accretion process.  In the rest of this chapter I will review some of the most recent highlights from this field.

\section{Matching simulations with observations}
\label{sec:matching}

One of the most exciting recent trends has been a concerted effort by various collaborations to make direct connections between very sophisticated simulations and observations.  Of course, observers have been clamoring for this sort of comparison for years!

Perhaps the first serious attempt at this was presented in \citet{schnittman_06}.  Schnittman produced a simulation similar to those in \citet{devilliers_03} and coupled it with a ray-tracing and radiative transfer code to produce ``images'' of what the simulated disk would look like to a distant observer.  By creating images from many time dumps in the simulation, Schnittman was able to create light curves, which were then analyzed for variability properties much the way real light curves are.  

\begin{figure}
\begin{center}
  \includegraphics[width=0.7\textwidth]{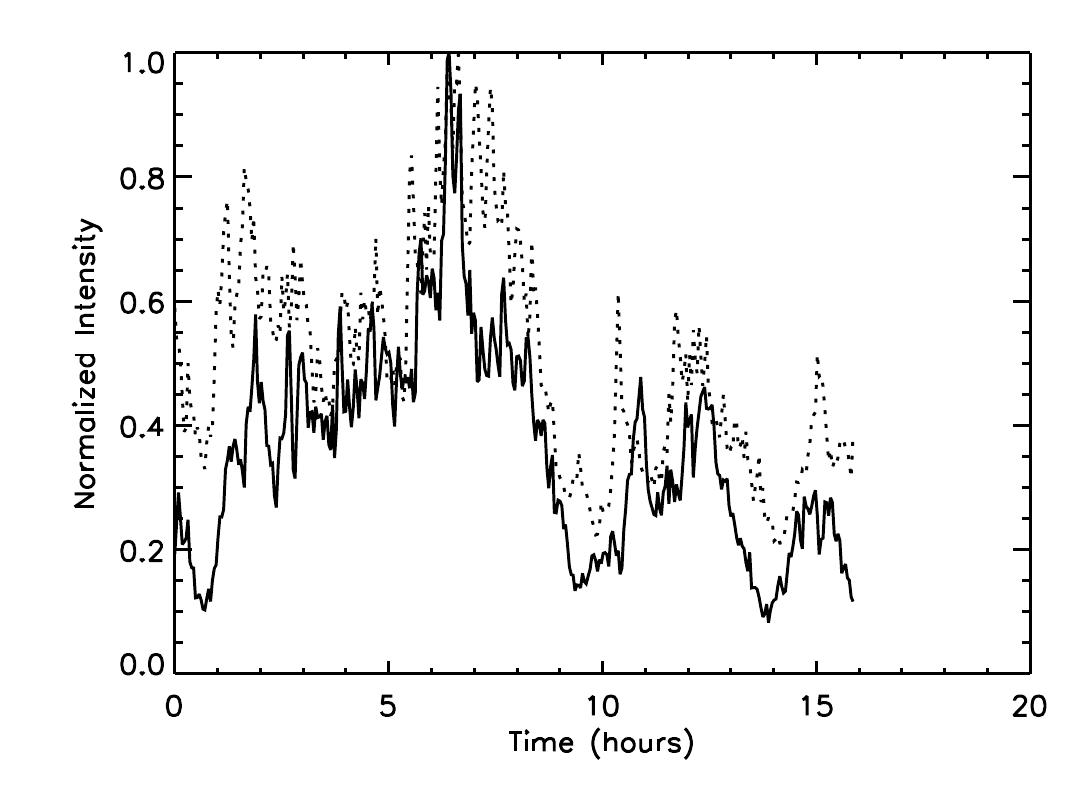}
\caption{Model light curve at 0.4 mm (solid) and accretion rate (dotted) at the inner boundary of a simulation from \citet{mckinney_09}.  Both quantities are scaled to their maximum value.  These are compared to data from Sgr A* in \citet{dexter_10}.}
\label{fig:lightcurve} 
\end{center}
\end{figure}

Following that same prescription, a number of groups have now presented results coupling general relativistic MHD (GRMHD) simulations with radiative modeling and ray-tracing codes \citep[e.g.][]{noble_07,moscibrodzka_09,dexter_09,dexter_10}.  More recent models have even included polarization measurements \citep{shcherbakov_12}.  This approach is most applicable to very low-luminosity systems, such as Sgr A* and M87.  A sample light curve for Sgr A* covering a 16-hour window is shown in Figure \ref{fig:lightcurve}.  In the case of M87, modeling has focused on accounting for the prominent jet in that system \citep{moscibrodzka_11,dexter_12}.

\begin{figure}
\begin{center}
  \includegraphics[width=0.7\textwidth]{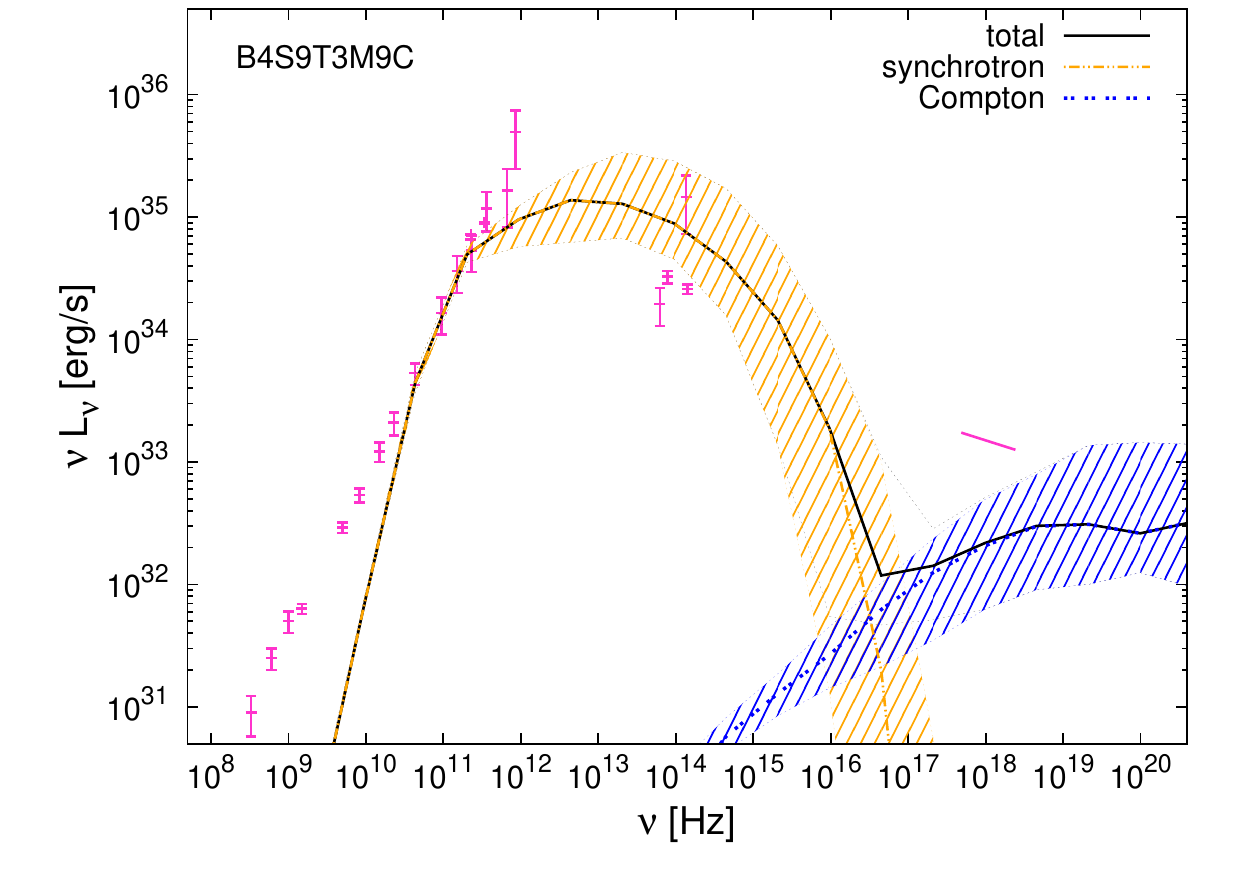}
\caption{Synthetic broadband spectrum created from one of the simulations presented in \citet{dibi_12}.  The pink points represent a compilation of Sgr A* observations.  Figure from \citet{drappeau_13}.}
\label{fig:spectrum} 
\end{center}
\end{figure}

Along with modeling light curves and variability, this approach can also be used to create synthetic broadband spectra from simulations \citep[e.g.][]{moscibrodzka_09,drappeau_13}, which can be compared with modern multi-wavelength observing campaigns (see Chapter 3.1).  This is very useful for connecting different components of the spectra to different regions of the simulation domain.  For example, Figure \ref{fig:spectrum} shows that the sub-mm bump in Sgr A* is well represented by emission from relatively cool, high-density gas orbiting close to the black hole, while the X-ray emission seems to come from Comptonization by very hot electrons in the highly magnetized regions of the black hole magnetosphere or base of the jet.

\section{Thermodynamics of simulations}
\label{sec:thermo}

As important as the radiative modeling of simulations described in Section \ref{sec:matching} has been, its application is very limited.  This is because, in most cases, the radiative modeling has been done after the fact; it was not included in the simulations themselves.  Therefore, the gas in the accretion disk was not allowed to respond thermodynamically to the cooling.  This calls into question how much the structure obtained from the simulation reflects the true structure of the disk.  Fortunately, various groups are beginning to work on treating the thermodynamics of accretion disks within the numerical simulations with greater fidelity.  Thus far, two approaches have principally been explored: 1) {\it ad hoc} cooling prescriptions used to artificially create {\it optically thick, geometrically thin} disks and 2) fully self-consistent treatments of radiative cooling for {\it optically thin, geometrically thick} disks.  We review each of these in the next 2 sections.

\subsection{Geometrically thin disks}
\label{sec:thin}

\begin{figure}
\begin{center}
  \includegraphics[width=0.7\textwidth]{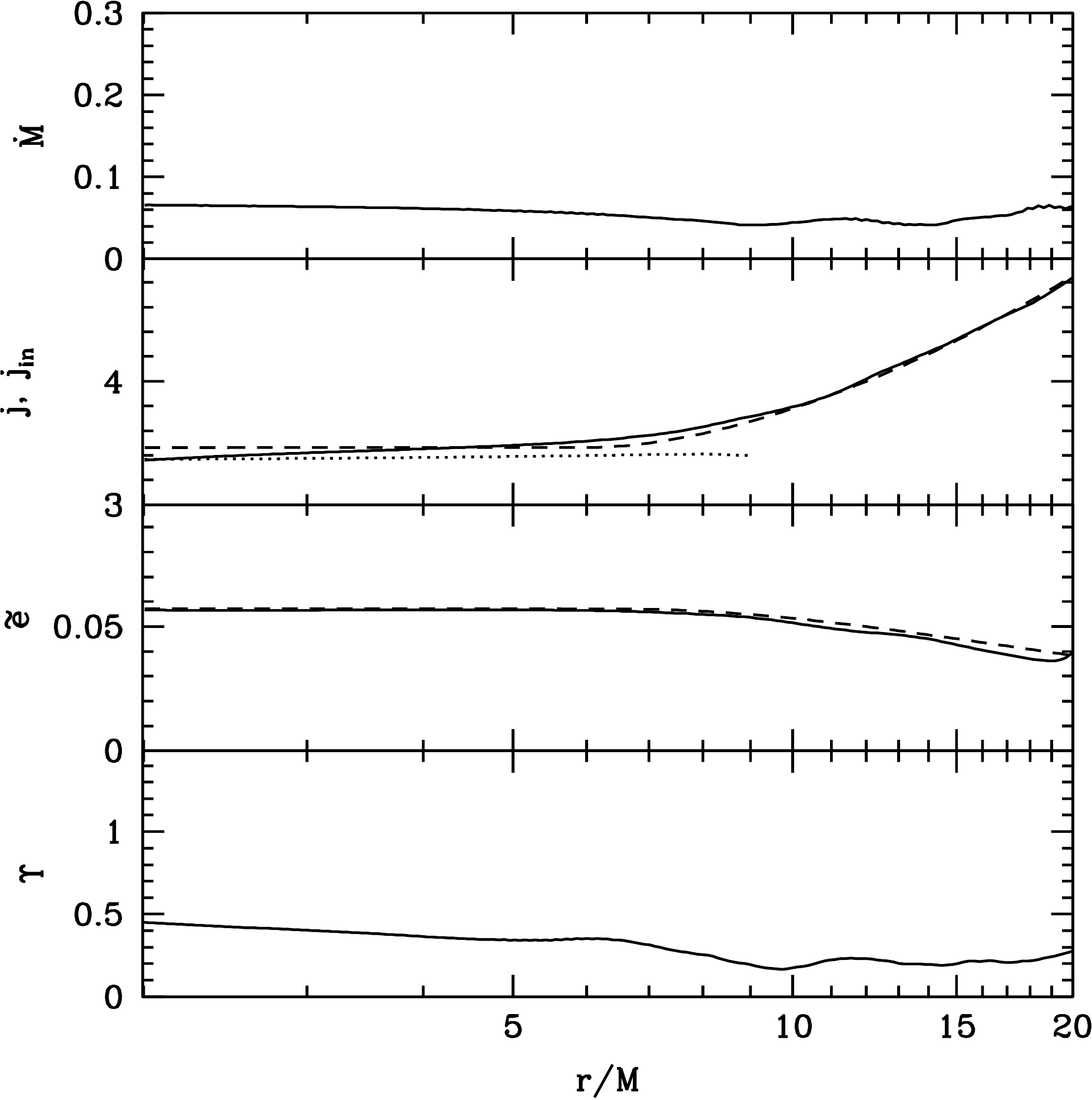}
\caption{Various fluxes as functions of radius for a numerical Novikov--Thorne disk simulation. \emph{Top:} Mass accretion rate. \emph{Second panel:} Accreted specific angular momentum.  \emph{Solid} line is simulation data; \emph{dashed} line gives Novikov--Thorne solution; \emph{dotted} line is ISCO value. Note that the specific angular momentum does not drop significantly inside the ISCO. \emph{Third panel:} The ``nominal'' efficiency, which is the total loss of specific energy from the fluid. \emph{Bottom panel:} Specific magnetic flux. The near constancy of this quantity inside the ISCO is an indication that magnetic stresses are not significant in this region. Figure from~\citet{penna_10}.}
\label{fig:ISCO} 
\end{center}
\end{figure}

For the {\it ad hoc} cooling prescription, cooling is assumed to equal heating (approximately) everywhere locally in the disk.  Since this is the same assumption as is made in the Shakura-Sunyaev \citep{shakura_73} and Novikov-Thorne \citep{novikov_73} disk models, this approach has proven quite useful in testing the key assumptions inherent in these models \citep[e.g][]{shafee_08,noble_09,penna_10,noble_10}.  In particular, these simulations have been useful for testing the assumption that the stress within the disk goes to zero at the innermost stable circular orbit (ISCO).  A corollary to this is that the specific angular momentum of the gas must remain constant at its ISCO value inside this radius.  Both of these effects have been confirmed in simulations of sufficiently thin disks \citep{penna_10}, as shown in Figure \ref{fig:ISCO}.

\subsection{Self-consistent radiative cooling of optically thin disks}
\label{sec:cooling}

Another approach to treating the thermodynamics of accretion disks has been to include {\it physical} radiative cooling processes directly within the simulations.  So far there has been very limited work done on this for optically thick disks, but an optically-thin treatment was introduced in \citet{fragile_09}.  Similar to the after-the-fact radiative modeling described in Section \ref{sec:matching}, the optically-thin requirement restricts the applicability of this approach to relatively low luminosity systems, such as the Quiescent and Low/Hard states of black hole X-ray binaries.

\begin{figure}
\begin{center}
  \includegraphics[width=0.7\textwidth]{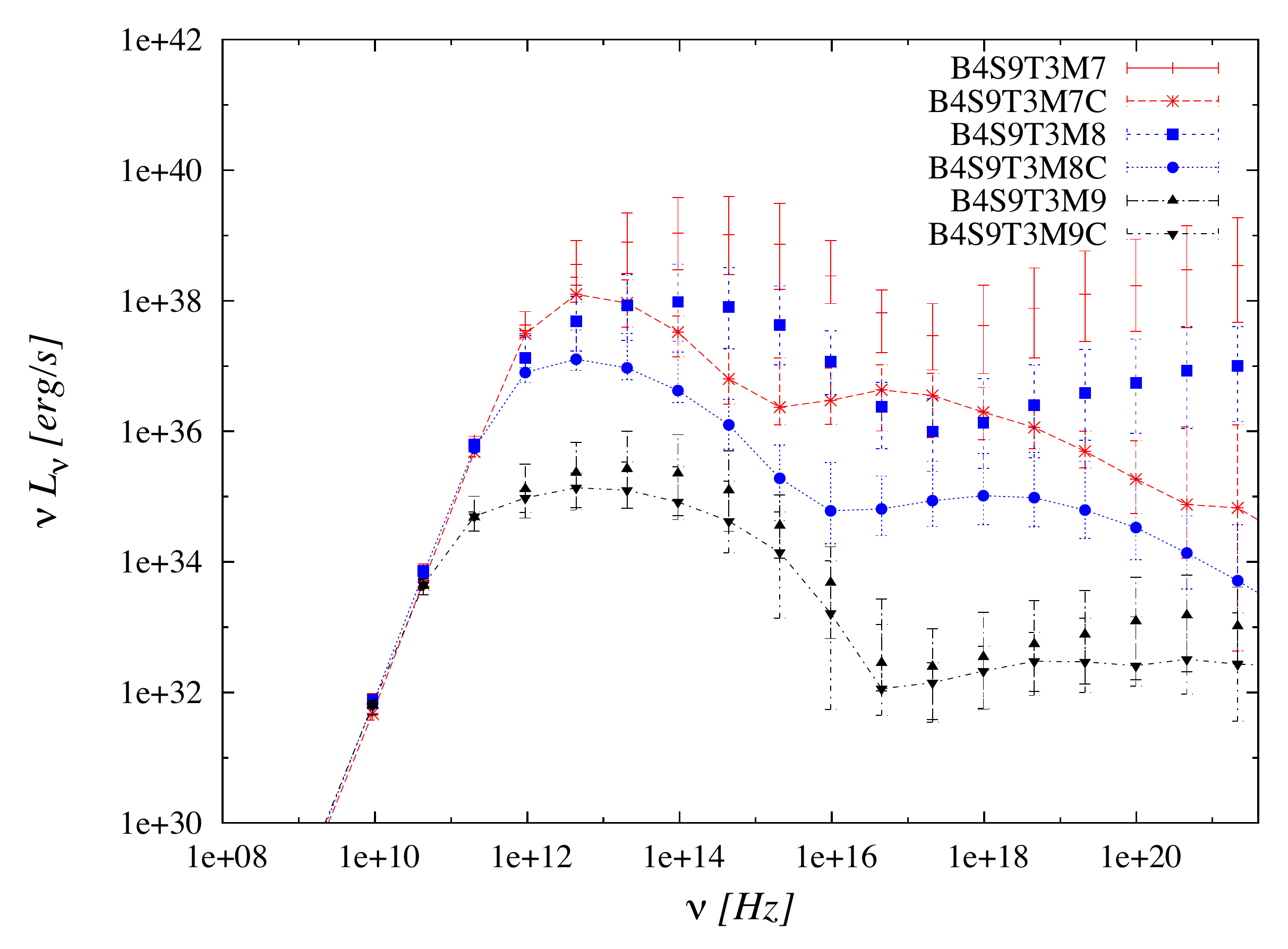}
\caption{Comparison of sample spectra generated for Sgr A* from numerical simulations at three different accretion rates: $10^{-9}$ ({\it black}), $10^{-8}$ ({\it blue}), and $10^{-7} M_\odot~\mathrm{yr}^{-1}$ ({\it red}).  For each accretion rate, two simulations are shown, one that includes cooling self-consistently (model names ending in ``C'') and one that does not.  The spectra begin to diverge noticeably at $\dot{M} \approx 10^{-8} M_\odot~\mathrm{yr}^{-1}$.  Figure from~\citet{dibi_12}.}
\label{fig:SgrA}
\end{center}
\end{figure}

Recently this approach has been applied to Sgr A* \citep{drappeau_13}, which turns out to be right on the boundary between where after-the-fact radiative modeling breaks down and a self-consistent treatment  becomes necessary \citep{dibi_12}.  Figure \ref{fig:SgrA} illustrates that this transition occurs right around an accretion rate of $\dot{M} \approx 10^{-8} M_\odot~\mathrm{yr}^{-1}$ for Sgr A*.

\section{Magnetic field topology}
\label{sec:field}

Another area where a lot of interesting new results have come out is in the study of how magnetic field topology and strength affect black hole accretion.  

\subsection{Jet power}
\label{sec:jet}

\begin{figure}
\begin{center}
  \includegraphics[width=\textwidth]{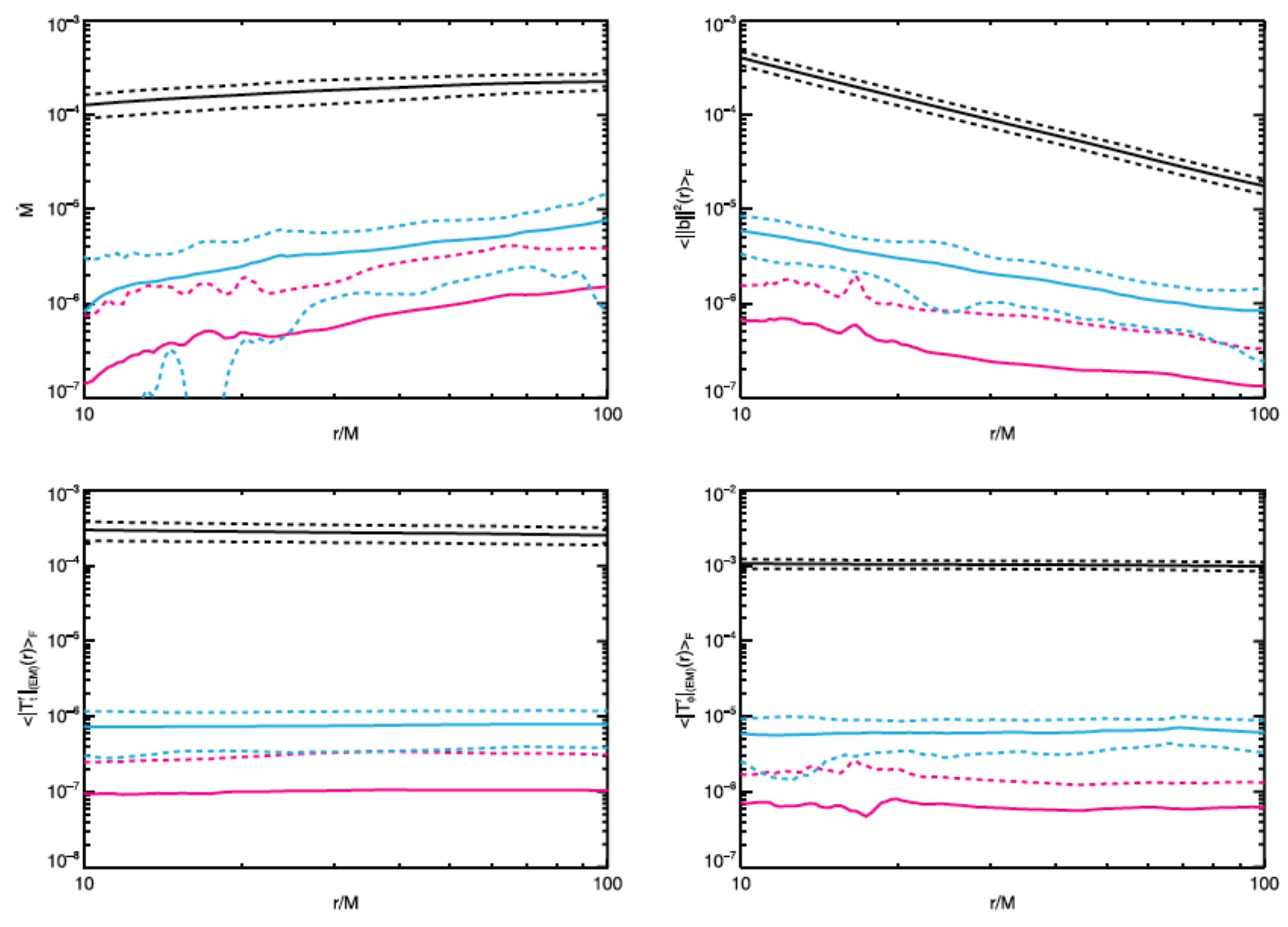}
\caption{Time-averaged shell integrals of data from unbound outflows as a function of radius for 3 models from \citet{beckwith_08}: a dipole magnetic field model ({\it black}), a quadrupole model ({\it cyan}), and a toroidal field model ({\it magenta}). Shown are mass outflow rate $\dot{M}$ ({\it top left}), magnetic field strength $\vert\vert b^2 \vert\vert$ ({\it top right}), electromagnetic energy flux $\vert T^r_t \vert_\mathrm{EM}$ ({\it bottom left}), and angular momentum flux $\vert T^r_\phi \vert_\mathrm{EM}$ ({\it bottom right}). Dashed lines show $\pm 1$ standard
deviation from the average.}
\label{fig:beckwith}
\end{center}
\end{figure}

Although there is now convincing evidence that the Blandford-Znajek mechanism \citep{blandford_77} works as predicted in powering jets \citep[e.g.][]{komissarov_01,mckinney_04}, one lingering question is still how the accretion process supplies the required poloidal flux onto the black hole.  Simulations have demonstrated that such field can, in many cases, be generated self-consistently within MRI-unstable disks \citep[e.g.][]{devilliers_05, hawley_06, mckinney_04, mckinney_06}.  However, this is strongly dependent on the initial magnetic field topology, as shown in \citet{beckwith_08}.  Figure \ref{fig:beckwith} nicely illustrates that when there is no net poloidal magnetic flux threading the inner disk, the magnetically-driven jet can be 2 orders of magnitude less energetic than when there is.  At this time it is unclear what the ``natural'' field topology would be, or even if there is one.

\subsection{Magnetically arrested accretion}
\label{sec:arrested}

\begin{figure}
\begin{center}
  \includegraphics[width=\textwidth]{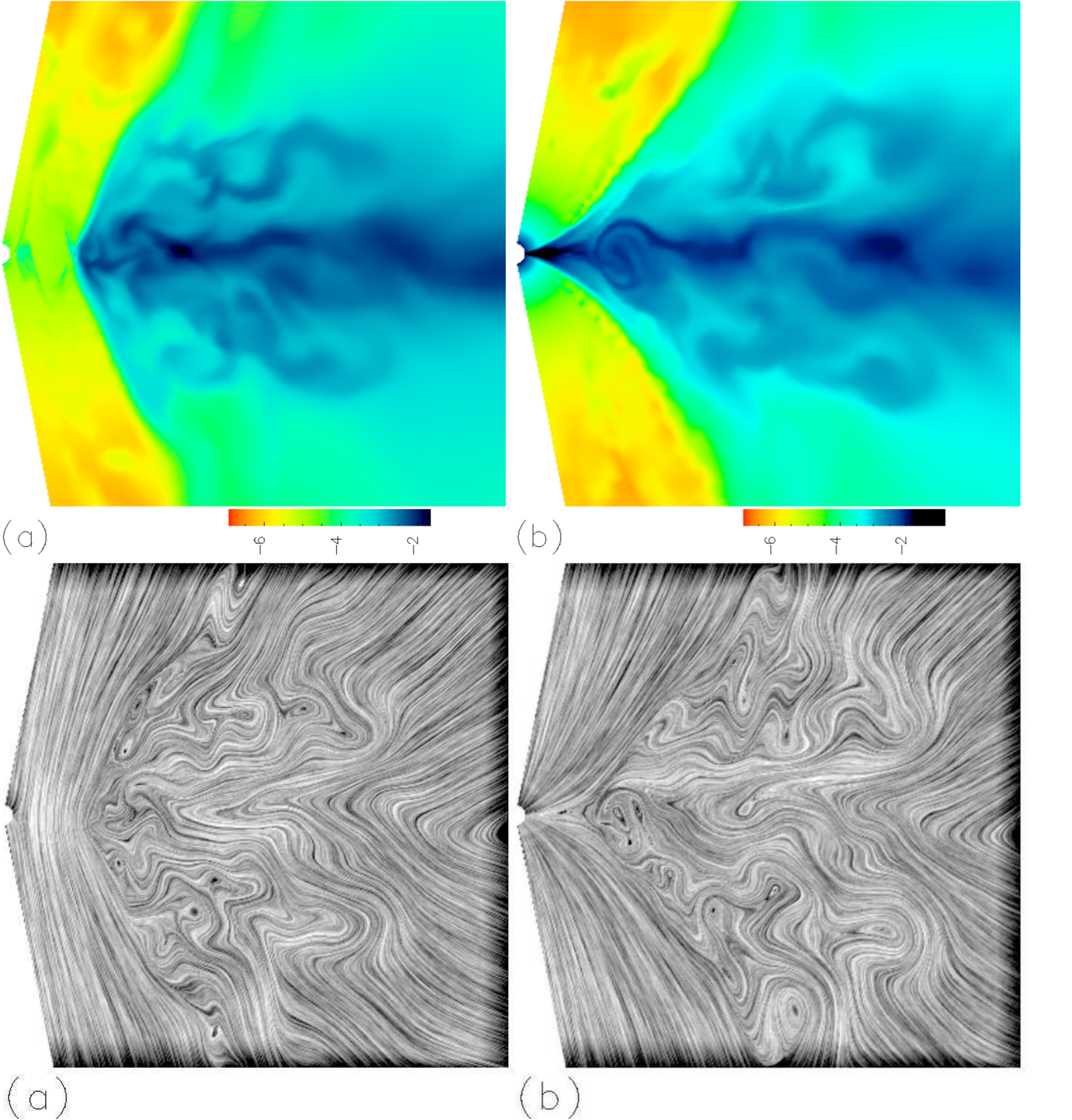}
\caption{Distributions of density in the meridional plane at different simulation times, showing a magnetically-arrested state (\emph{left}) and a non-arrested state (\emph{right}). \emph{Bottom:} Snapshot of magnetic field lines at the same simulation times. Figure from~\citet{igumenshchev_08}.}
\label{fig:arrest}
\end{center}
\end{figure}

Although strong poloidal magnetic fields are useful for driving powerful jets, they can also create interesting feedback affects on an accretion disk.  In the case where a black hole is able to accumulate field with a consistent net flux for an extended period of time, it is possible for the amassed field to eventually ``arrest'' the accretion process \citep{narayan_03}.  An example of an arrested state is shown in Figure \ref{fig:arrest}.

\begin{figure}
\begin{center}
  \includegraphics[width=0.7\textwidth]{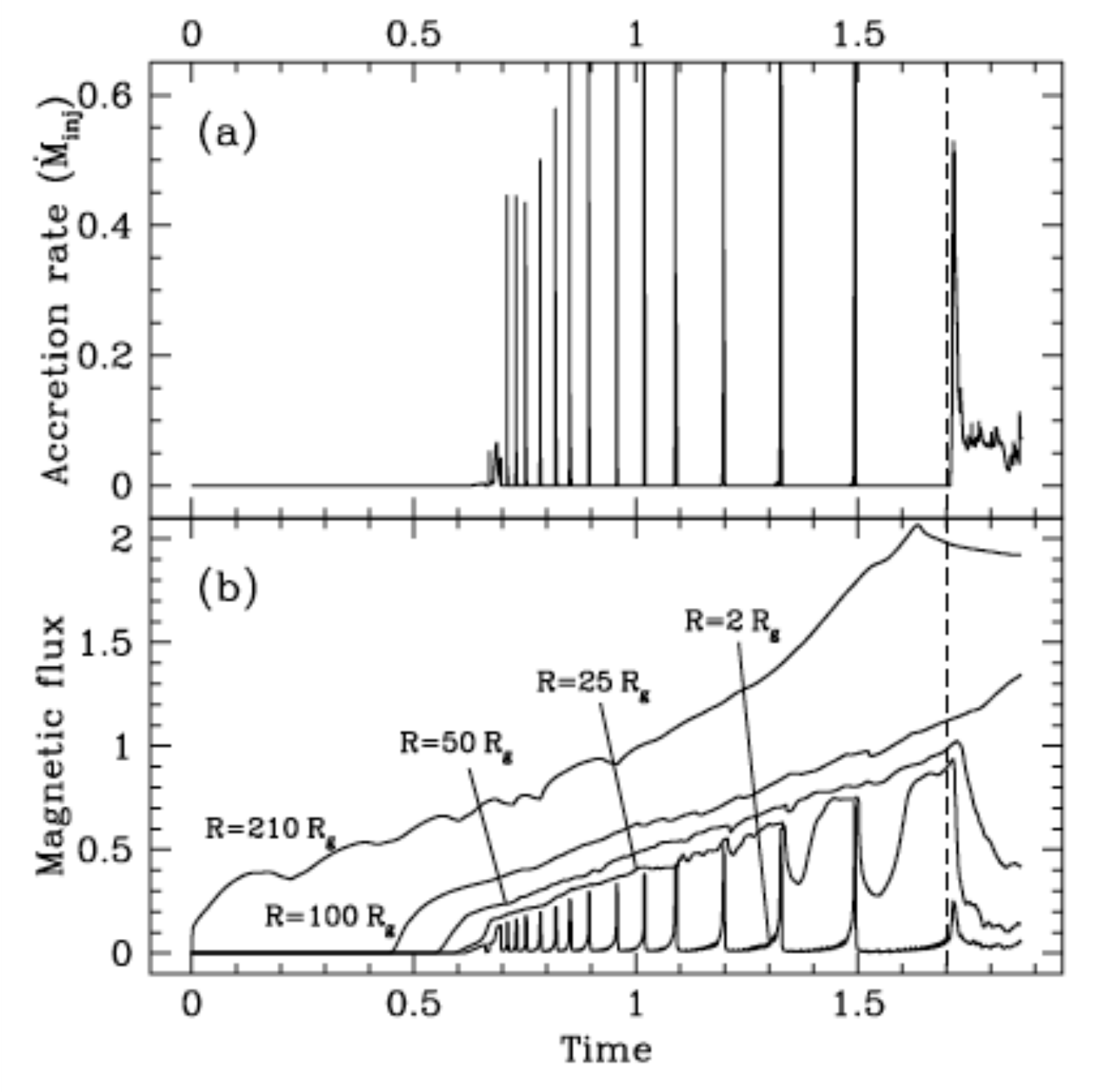}
\caption{Evolution of mass accretion rate and magnetic fluxes in 2D (axisymmetric) simulation of magnetically-arrested accretion. Accretion into the black hole begins at $t \approx 1.3$. Starting from $t \approx 1.4$, a pattern of cyclic accretion develops (seen as a sequence of spikes). Figure from~\citet{igumenshchev_08}.}
\label{fig:arrested_mdot}
\end{center}
\end{figure}

In a two-dimensional simulation where plasma with a constant net flux is fed in from the outer boundary, a limit-cycle behavior can set in, where the mass accretion rate varies by many orders of magnitude between the arrested and non-arrested states.  Figure \ref{fig:arrested_mdot} provides an example of the resulting mass accretion history.  It is straightforward to show that the interval, $\Delta t$, between each non-arrested phase in this scenario grows with time according to 
\begin{equation}
\Delta t \sim \frac{2 v_r B_z^2}{\rho} t^2~,
\end{equation}
where $v_r$ is the radial infall velocity of the gas, $B_z$ is the strength of the magnetic field, and $\rho$ is the density of the gas.

In three-dimensions, the magnetic fields are no longer able to perfectly arrest the in-falling gas because of a ``magnetic'' Rayleigh-Taylor effect.  Basically, as low density, highly magnetized gas tries to support higher density gas in a gravitational potential, it becomes unstable to an interchange of the low- and high-density materials.  Indeed, such a magnetic Rayleigh-Taylor effect has been seen in recent simulations by \citet{igumenshchev_08,tchekhovskoy_11,mckinney_12}.  
\begin{figure}
\begin{center}
  \includegraphics[width=\textwidth]{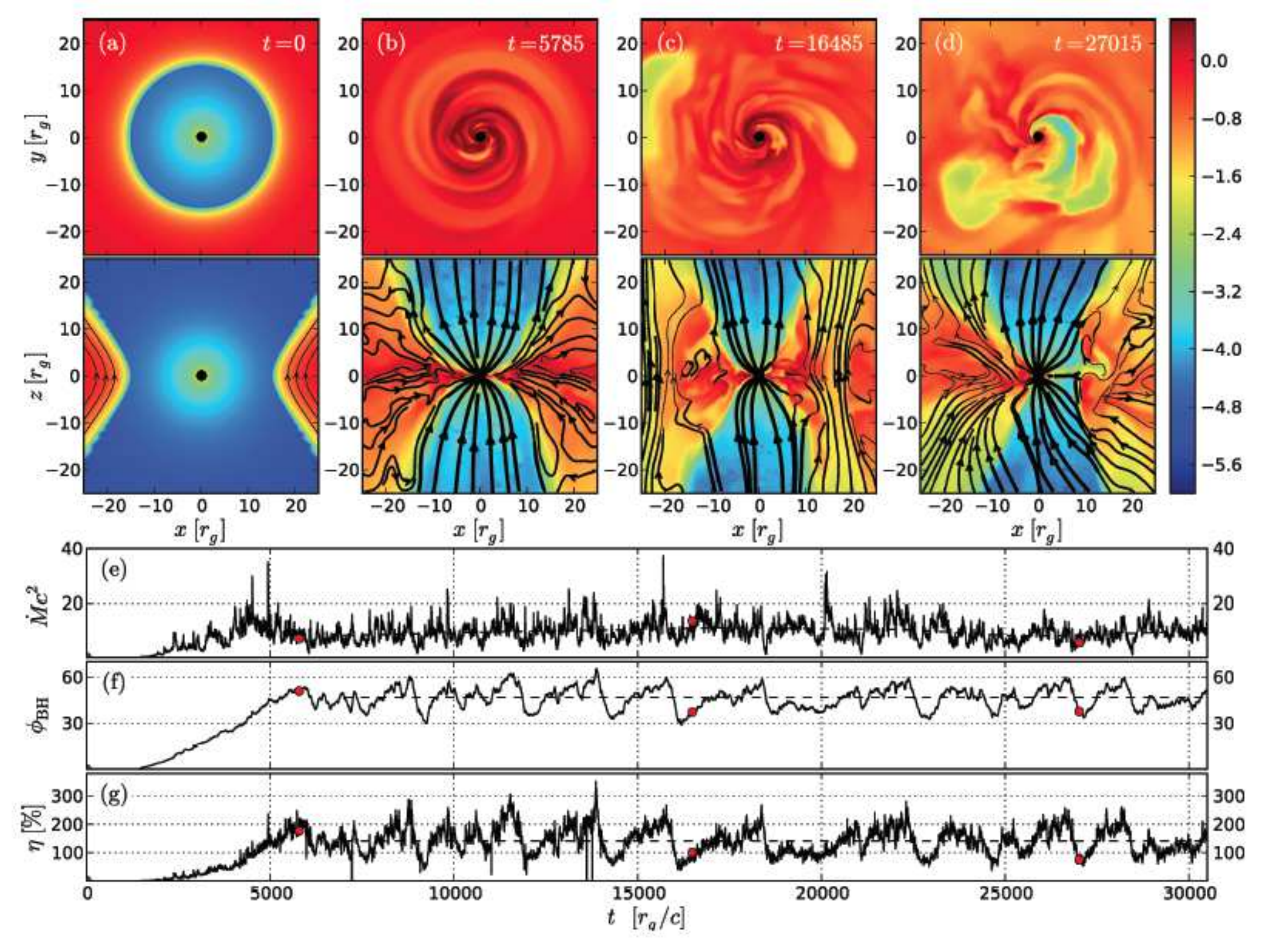}
\caption{(Panels a-d): the top and bottom rows show, respectively, equatorial ($z = 0$) and meridional ($y = 0$) snapshots of the gas density of a magnetically arrested flow in 3D.  Black lines show field lines in the image plane. (Panel e): time evolution of the mass accretion rate. (Panel f): time evolution of the large-scale magnetic flux threading the BH horizon. (Panel g): time evolution of the energy outflow efficiency $\eta$. Figure from~\citet{tchekhovskoy_11}.}
\label{fig:tchekhovskoy}
\end{center}
\end{figure}

The results of \citet{tchekhovskoy_11} are important for another reason.  These were the first simulations to demonstrate a jet efficiency $\eta = (\dot{M} - \dot{E})/\langle \dot{M} \rangle$ greater than unity.  Since the efficiency measures the amount of energy extracted by the jet, normalized by the amount of energy made available via accretion, a value $\eta > 1$ indicates more energy is being extracted than is being supplied by accretion.  This is only possible if some other source of energy is being tapped -- in this case the rotational energy of the black hole.  This was the first demonstration that a Blandford-Znajek \citep{blandford_77} process {\it must} be at work in driving these simulated jets.

\section{Tilted disks}
\label{sec:tilt}

There is observational evidence that several black-hole X-ray binaries (BHBs), e.g. GRO J1655-40 \citep{orosz_97}, V4641 Sgr \citep{miller_02} and GX 339-4 \citep{miller_09}, and active galactic nuclei (AGN), e.g. NGC 3079 \citep{kondratko_05}, NGC 1068 \citep{caproni_06}, and NGC 4258 \citep{caproni_07}, may have accretion disks that are tilted with respect to the symmetry plane of their central black hole spacetimes. There are also compelling theoretical arguments that many black hole accretion disks should be tilted \citep{fragile_01,maccarone_02}.  This applies to both stellar mass black holes, which can become tilted through asymmetric supernovae kicks \citep{fragos_10} or binary captures and will remain tilted throughout their accretion histories, and to supermassive black holes in galactic centers, which will likely be tilted for some period of time after every major merger event \citep{kinney_00}.

Close to the black hole, tilted disks may align with the symmetry plane of the black hole, either through the Bardeen-Petterson effect \citep{bardeen_75} in geometrically thin disks or through the magneto-spin alignment effect \citep{mckinney_13} in the case of geometrically thick, magnetically-choked accretion \citep{mckinney_12}.  However, for weakly magnetized, moderately thick disks ($H/r \gtrsim 0.1$), no alignment is observed \citep{fragile_07}.  In such cases, there are many observational consequences to consider.

\subsection{Tilted disks and spin}

\begin{figure}
\begin{center}
  \includegraphics[width=0.7\textwidth]{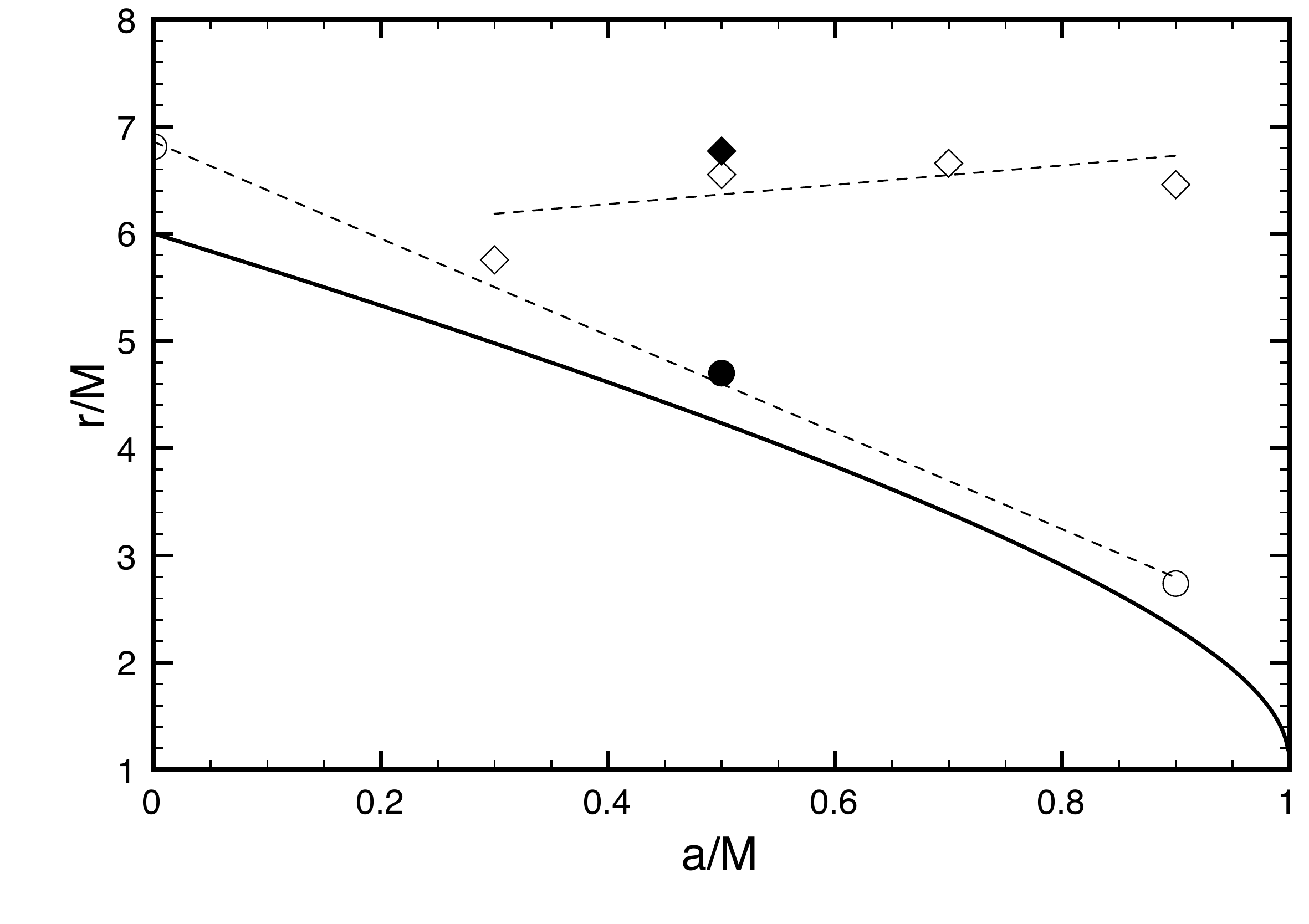}
\caption{Plot of the effective inner radius $r_\mathrm{in}$ of simulated untilted ({\em circles}) and tilted ({\em diamonds}) accretion disks as a function of black-hole spin using a surface density measure $\Sigma(r_\mathrm{in}) = \Sigma_\mathrm{max}/3e$. The {\em solid} line is the ISCO radius. Figure from~\citet{fragile_09b}.}
\label{fig:fragile09}
\end{center}
\end{figure}

Chapter 4.3 of this book discusses the two primary methods for estimating the spins of black holes: continuum-fitting and reflection-line modeling.  Both rely on an assumed monotonic relation between the inner edge of the accretion disk (assumed to coincide with the radius of the ISCO) and black hole spin, $a$.  This is because what both methods actually measure is the effective inner radius of the accretion disk, $r_\mathrm{in}$.  One problem with this is that it has been shown \citep{fragile_09b,dexter_11} that tilted disks do not follow such a monotonic behavior, at least not for disks that are not exceptionally geometrically thin. Figure~\ref{fig:fragile09} shows an example of the difference between how $r_\mathrm{in}$ depends on $a$ for untilted and tilted simulated disks.  Similar behavior has been confirmed using both dynamical \citep{fragile_09b} and radiative \citep{dexter_11} measures of $r_\mathrm{in}$. The implication is that spin can only be reliably inferred in cases where the inclination of the inner accretion disk can be independently determined, such as by modeling jet kinematics \citep{steiner_12}.

\subsection{Tilted disks and Sgr A* spectral fitting}

\begin{figure}
\begin{center}
  \includegraphics[width=\textwidth]{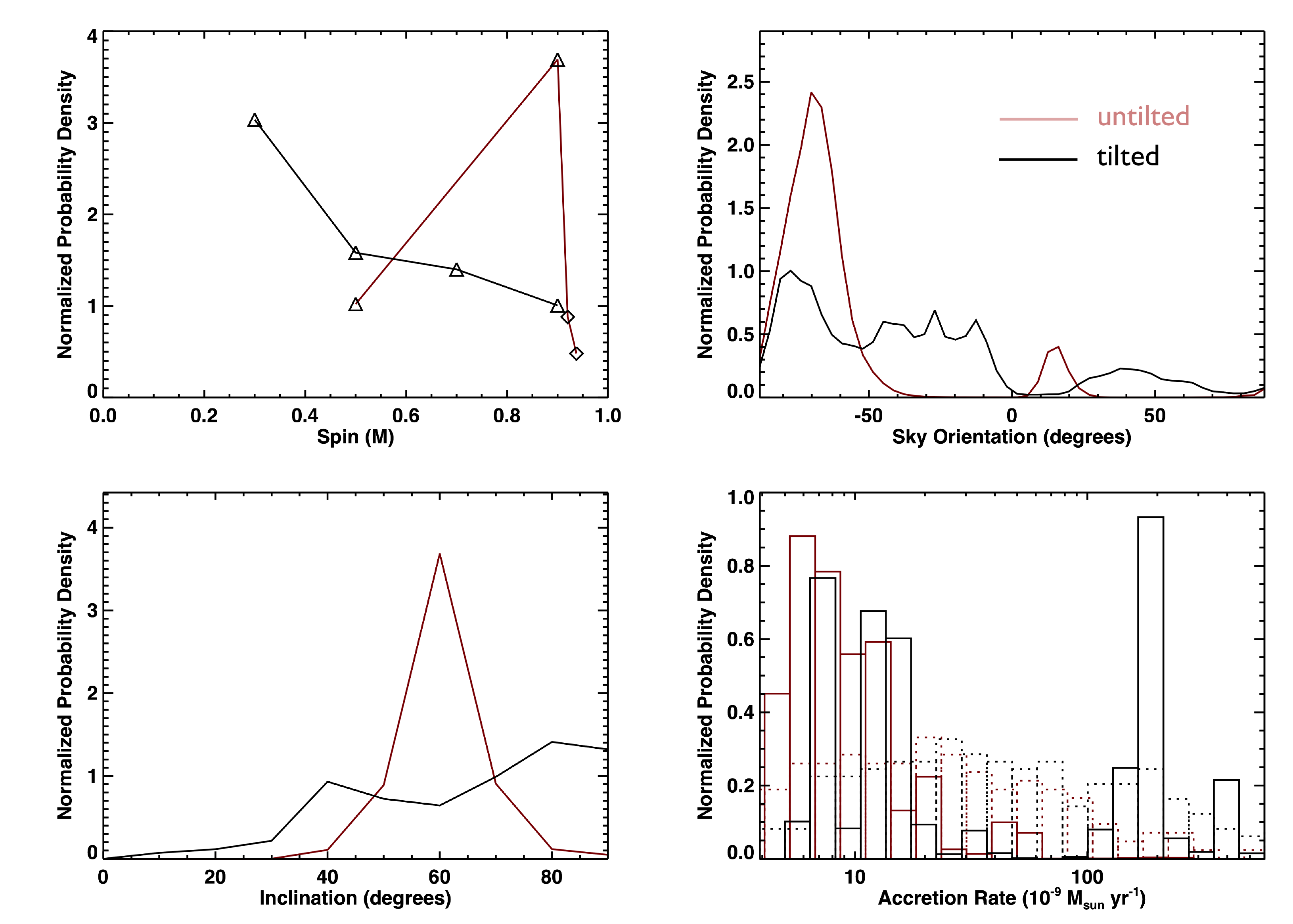}
\caption{Normalized probability distributions as a function of black hole spin (top left), sky orientation (top right), inclination (bottom left), and accretion rate (bottom right) for untilted (red) and tilted (black) simulations. Figure adapted from \citet{dexter_10} and \citet{dexter_13}.}
\label{fig:dexter12}
\end{center}
\end{figure}

For geometrically thin, Shakura-Sunyaev type accretion disks, the Bardeen-Petterson effect \citep{bardeen_75} may allow the inner region of the accretion disk to align with the symmetry plane of the black hole, perhaps alleviating concerns about measuring $a$, at least for systems in the proper state (``Soft'' or ``Thermally dominant'') and luminosity range $L \le 0.3 L_\mathrm{Edd}$, where $L_\mathrm{Edd}$ is the Eddington luminosity.  Extremely low luminosity systems, though, such as Sgr A*, do not experience Bardeen-Petterson alignment.  Further, for a system like Sgr A* that is presumed to be fed by winds from massive stars orbiting in the galactic center, there is no reason to expect the accretion flow to be aligned with the black hole spin axis.  Therefore, a tilted configuration should be expected. In light of this, \citet{dexter_13} presented an initial comparison of the effect of tilt on spectral fitting of Sgr A*.  Figure \ref{fig:dexter12} gives one illustration of how important this effect is; it shows how the probability density distribution of four observables change if one simply accounts for the two extra degrees of freedom introduced by even a modestly tilted disk.  The take away point should be clear -- ignoring tilt artificially constrains these fit parameters!

\subsection{Tilted disks and strong shocks}

\begin{figure}
\begin{center}
  \includegraphics[width=0.8\textwidth]{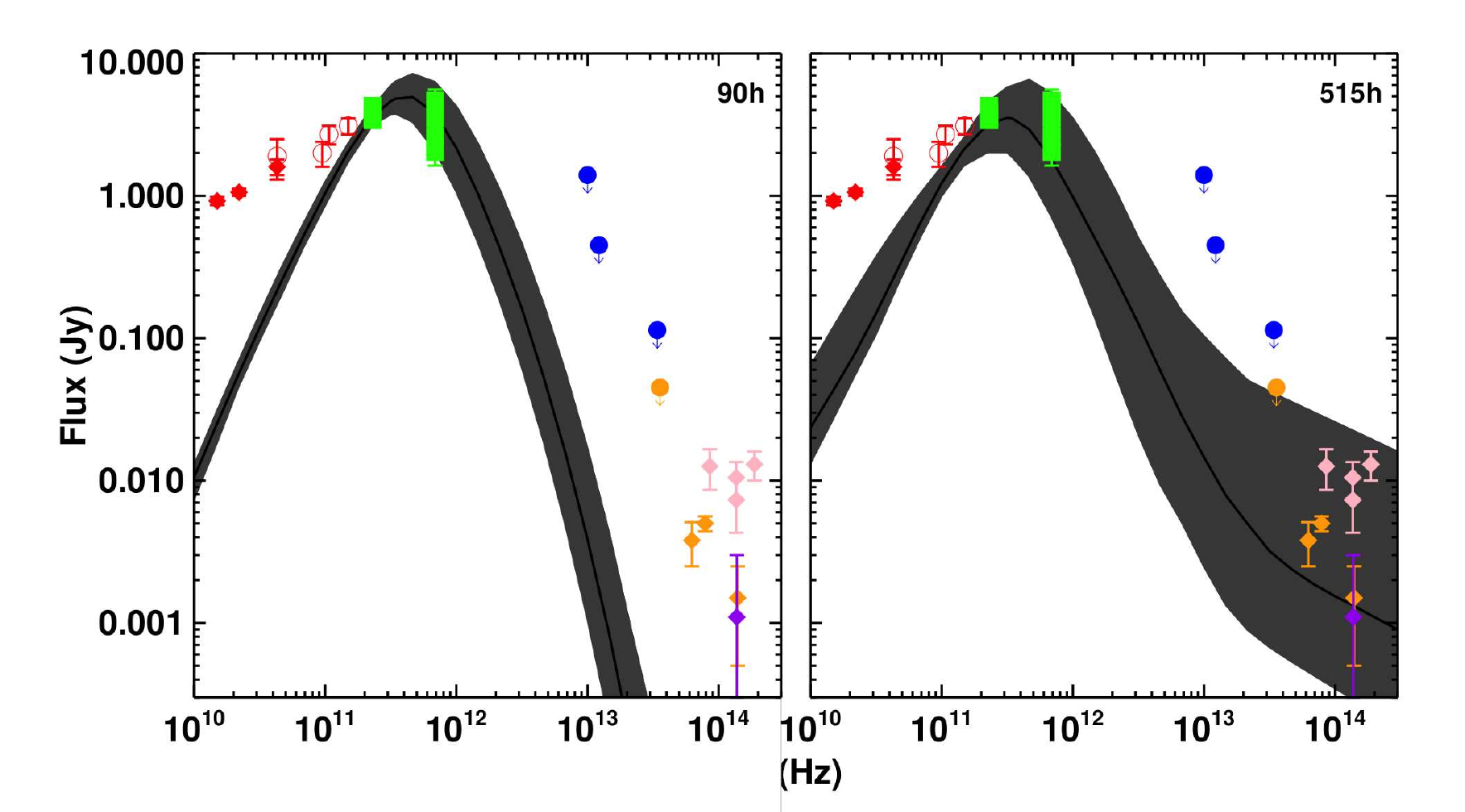}
\caption{Spectra from untilted (left) and tilted (right) simulations. Symbols represent Sgr A* data. In both cases the spectra are fit to the green sub-mm points. In the tilted simulations, multiple electron populations, some heated by shocks associate with the tilt, are present and can naturally produce the observed NIR emission, which is underproduced by $\approx 2$ orders of magnitude in comparable untilted simulations.  Figure from \citet{dexter_13}.}
\label{fig:tilted_spectrum}
\end{center}
\end{figure}

\begin{figure}
\begin{center}
  \includegraphics[width=0.7\textwidth]{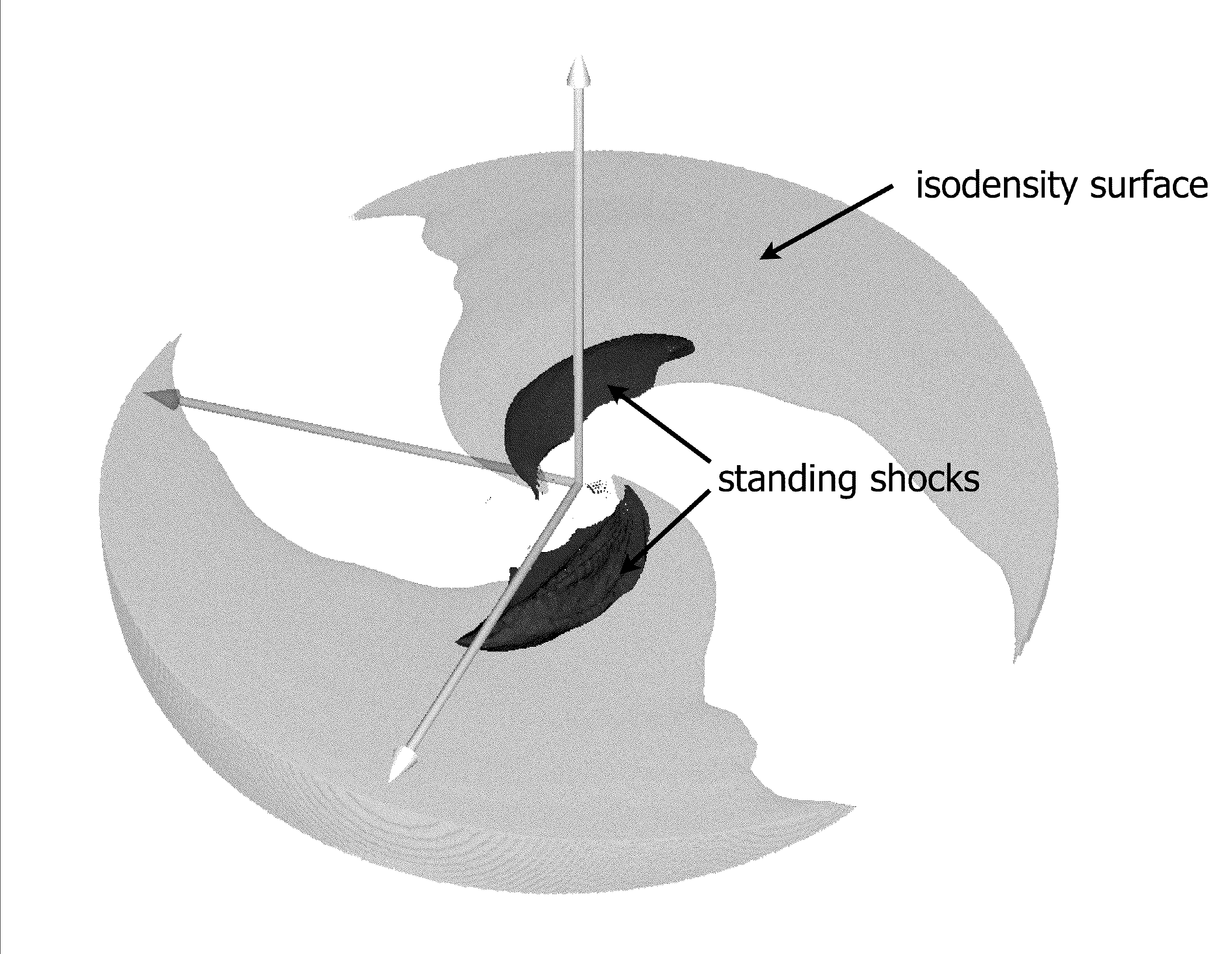}
\caption{Three-dimensional contours of density (semi-transparent gray) and Lagrangian specific entropy generation rate in arbitrary units (solid gray), indicating the shocks.  Figure from \citet{henisey_12}.}
\label{fig:henisey12}
\end{center}
\end{figure}

One remarkable outcome of considering tilt in fitting the spectral data for Sgr A* is that tilted simulations seem able to naturally resolve a problem that had plagued earlier studies.  Spectra produced from untilted simulations of Sgr A* have always yielded a deficit of flux in the near-infrared compared to what is observed.  For untilted simulations, this can only be rectified by invoking additional spectral components beyond those that naturally arise from the simulations.  Tilted simulations, though, produce a sufficient population of hot electrons, {\it without any additional assumptions}, to produce the observed near-infrared flux (see comparison in Figure \ref{fig:tilted_spectrum}).  They are able to do this because of another unique feature of tilted disks: the presence of standing shocks near the line-of-nodes at small radii \citep{fragile_08}.  These shocks are a result of epicyclic driving due to unbalanced pressure gradients in tilted disks leading to a crowding of orbits near their respective apocenters.  Figure \ref{fig:henisey12} shows the orientation of these shocks in relation to the rest of the inner accretion flow.

\subsection{Tilted disks - GRMHD vs. SPH}

A worthwhile future direction to pursue in this area would be a robust comparison of tilted disk simulations using both GRMHD and smoothed-particle hydrodynamics (SPH) numerical methods.  The GRMHD simulations \citep[e.g.][]{fragile_07,fragile_09c,mckinney_13} enjoy the advantage of being ``first principles'' calculations, since they include all of the relevant physics, whereas the SPH simulations \citep[e.g.][]{nelson_00,lodato_07,lodato_10,nixon_12} enjoy the advantage of being more computationally efficient, though they make certain assumptions about the form of the ``viscosity'' in the disk.  Thus far, the GRMHD and SPH communities have proceeded separately in their studies of tilted accretion disks, and it has yet to be demonstrated that the two methods yield equivalent results.  This would seem to be a relatively straightforward and important thing to check.

\section{Future direction - radiation MHD}

\begin{figure}
\begin{center}
  \includegraphics[width=\textwidth]{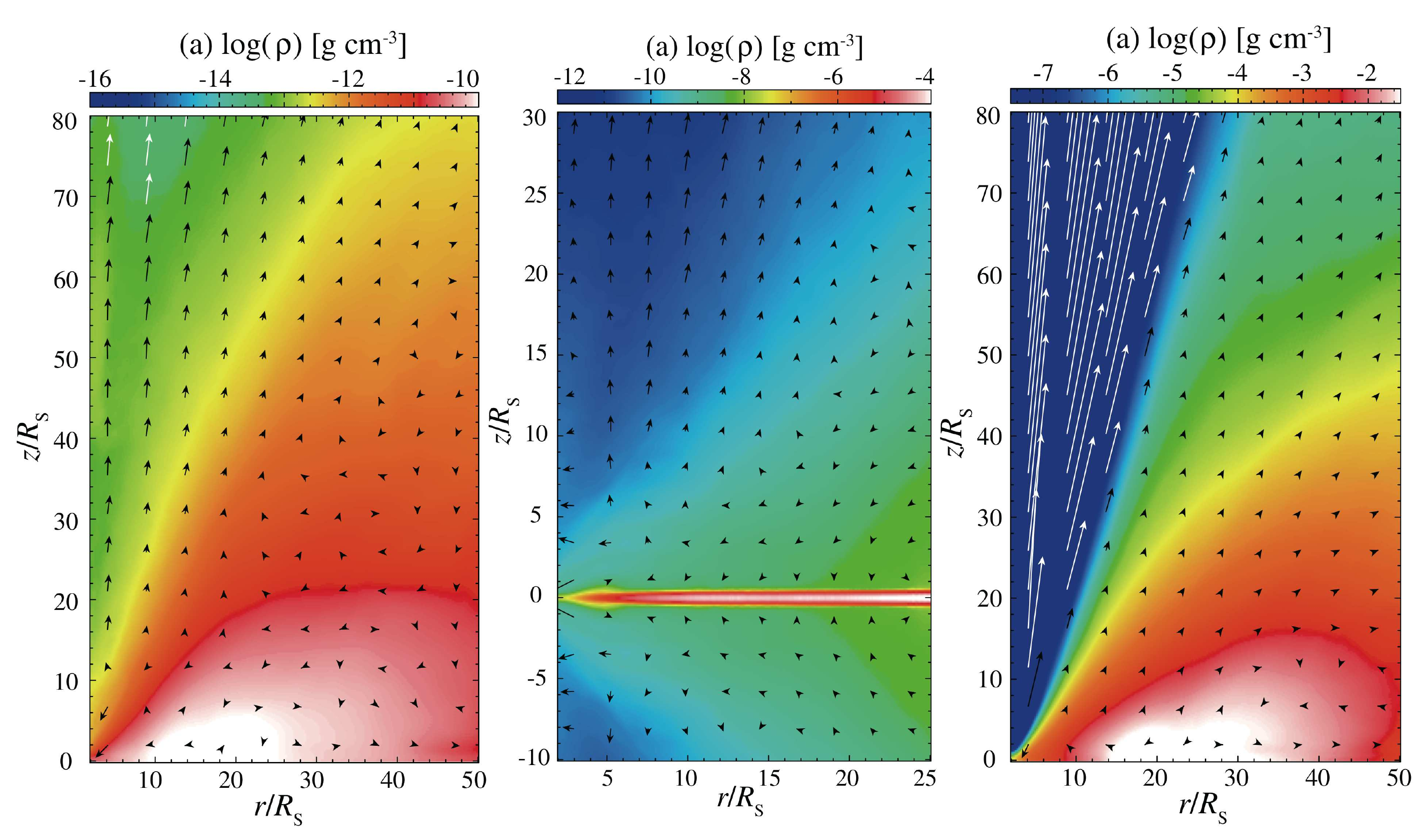}
\caption{Gas density overlaid with velocity vectors from three different radiation MHD simulations of a black hole accretion disk, probing luminosities of $L \sim 10^{-8}$ (left), $10^{-4}$(center), and $1 L_\mathrm{Edd}$ (right).  Figure adapted from \citet{ohsuga_11}.}
\label{fig:ohsuga11}
\end{center}
\end{figure}

A few years ago, it might have been very ambitious to claim that researchers would soon be able to perform global radiation MHD simulations of black hole accretion disks, yet a lot has happened over that time, so that now it is no longer a prediction but a reality.  In the realm of Newtonian simulations, a marvelous study was published by \citet{ohsuga_11}, showing global (though two-dimensional) radiation MHD simulations of accretion onto a black hole in three different accretion regimes: $L \ll L_\mathrm{Edd}$, $L < L_\mathrm{Edd}$, and $L \gtrsim L_\mathrm{Edd}$.  The remarkably different behavior of the disk in each of the simulations (illustrated in Figure \ref{fig:ohsuga11}) is testament to how rich this field promises to be as more groups join this line of research.  The specifics of this work are discussed more in Chapter 5.3.

\begin{figure}
\begin{center}
  \includegraphics[width=\textwidth]{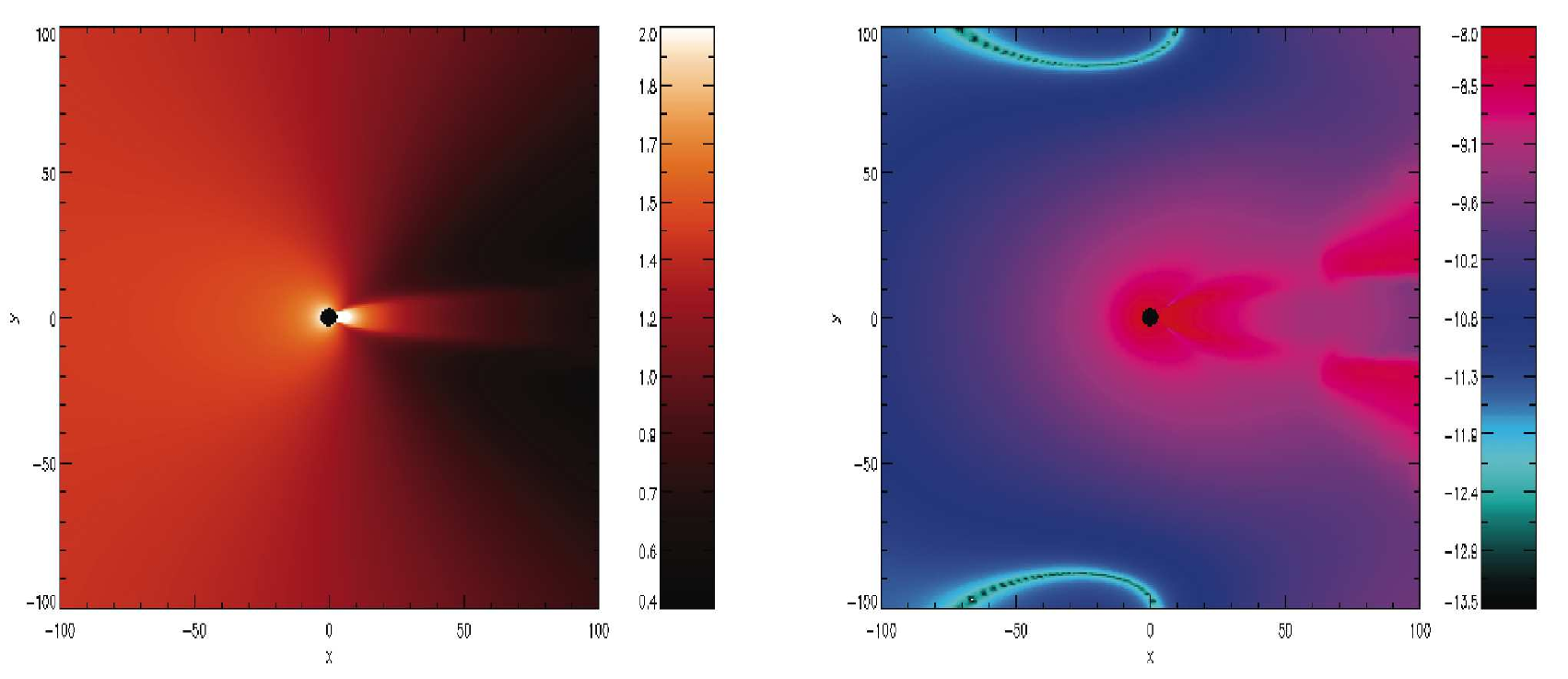}
\caption{Logarithm of the optical depth (left) and radiative flux (right) of a black hole accreting from a wind passing from left to right.  Figure from \citet{zanotti_11}.}
\label{fig:zanotti11}
\end{center}
\end{figure}

The other big thing to happen (mostly) within the past year is that a number of groups have now tackled, for the first time, the challenge of developing codes for {\it relativistic} radiation MHD in black hole environments \citep{farris_08,zanotti_11,roedig_12,fragile_12,sadowski_13,takahashi_13}.  So far none of these groups have gotten to the point of simulating accretion disks in the way \citet{ohsuga_11} did (they are still mostly at the stage of code tests and simple one- and two-dimensional problems), but with so many groups joining the chase, one can surely expect rapid progress in the near future.  One result of some astrophysical interest is the study of Bondi-Hoyle (wind) accretion onto a black hole, including the effects of radiation \citep{zanotti_11,roedig_12} (see Figure \ref{fig:zanotti11}).

\begin{acknowledgements}
Work presented in this chapter was supported in part by a High-Performance Computing grant from Oak Ridge Associated Universities/Oak Ridge National Laboratory and by the National Science Foundation under grant no. AST-0807385.
\end{acknowledgements}


\end{document}